# Single Photon Emission from Plasma Treated 2D Hexagonal Boron Nitride

*Zai-Quan Xu[†][*], Christopher Elbadawi[†], Toan Trong Tran[†], Mehran Kianinia[†], Xiuling Li[‡], Daobin Liu[^], Timothy B. Hoffman[§], Minh Nguyen[†], Sejeong Kim[†], James H. Edgar[§], Xiaojun Wu[‡], Li Song[^], Sajid Ali[†], Mike Ford[†], Milos Toth[†]\* and Igor Aharonovich[†]\**

[†] School of Mathematical and Physical Sciences, Faculty of Science, University of Technology Sydney, Ultimo, 2007, New South Wales, Australia

[‡] School of Chemistry and Materials Sciences, CAS Key Lab of Materials for Energy Conversion, and CAS Center for Excellence in Nanoscience, Hefei National Laboratory of Physics at the Microscale, Synergetic Innovation of Quantum Information & Quantum Technology, University of Science and Technology of China, Hefei, Anhui 230026, China

[^] National Synchrotron Radiation Laboratory, CAS Center for Excellence in Nanoscience, University of Science and Technology of China, Hefei, Anhui 230029, China.

[§] Department of Chemical Engineering, Durland Hall, Kansas State University, Manhattan, KS 66506, USA

Corresponding Emails: Zaiquan.xu@uts.edu.au; Milos.Toth@uts.edu.au; Igor.Aharonovich@uts.edu.au

**Abstract**

Artificial atomic systems in solids are becoming increasingly important building blocks in quantum information processing and scalable quantum nanophotonic networks. Yet, synthesis of color centers that act as single photon emitters which are suitable for on-chip applications is still beyond reach. Here, we report a number of plasma and thermal annealing methods for the fabrication of emitters in tape-exfoliated hexagonal boron nitride (hBN) crystals. A two-step process comprised of Ar plasma etching and subsequent annealing in Ar is highly robust, and yields a seven-fold increase in the concentration of emitters in hBN. The initial plasma etching step generates emitters that suffer from blinking and bleaching, whereas the two-step process yields emitters that are photostable at room temperature and have an emission energy distribution that is red-shifted relative to that of pristine hBN. An analysis of emitters fabricated by a range of plasma and annealing treatments, combined with a theoretical investigation of point defects in hBN indicates that single photon emitters characterized by a high degree of photostability and emission wavelengths greater than ~700 nm are associated with defect complexes that contain oxygen. This is further confirmed by generating the emitters by annealing hBN in an oxidative atmosphere. Our findings advance present understanding of the structure of quantum emitter in hBN and enhance the nanofabrication toolkit that is needed to realize integrated quantum nanophotonics based on 2D materials.

**Introduction**

Two-dimensional (2D) materials such as graphene, hexagonal boron nitride (hBN) and transition metal chalcogenides are of great interest in photonics and optoelectronics due to their unique optoelectronic, layer-dependent properties.[1-5] In particular, hBN is a layered wide band gap semiconductor[6] with covalently bonded boron and nitrogen atoms arranged in a honeycomb structure.[7] Due to its hyperbolic properties, hBN crystals are frequently used as substrates to enhance electronic and optical properties of other 2D materials.[8] Recently, hBN has been shown to host ultra-bright, room temperature, polarized emitters of non-classical light, attributed to localized point defects.[7, 9-11, 12] The defects exhibit a broad range of emission energies and their structural origin is a matter of debate. While several methods have recently been explored, including laser ablation[12], ion/electron irradiation[9] and chemical etching[13], they have so far had marginal success and low emitter fabrication yields. Consequently, new fabrication techniques of photostable quantum emitters are needed to enable useful on-chip devices[14]. Here, we elucidate the role of oxygen in fluorescent properties of hBN and show that a two-step Ar plasma etching and annealing process is a robust, scalable method for the fabrication of photostable single photon emitters (SPEs) in hBN. The ability to engineer the emitters deterministically in specific hBN crystals opens new pathways to integrate them in quantum nanophotonic devices.[15]

**Results and Discussions**

We start by demonstrating controlled fabrication of single photon emitters in hBN by Ar plasma etching. Figures 1a and 1b show an optical microscope image of tape-exfoliated hBN crystals on a silicon substrate, and a schematic illustration of the plasma etch process. Figures 1c and 1d show normalized, panchromatic confocal PL maps acquired at room temperature from the same crystal before and after Ar plasma etching. Unless noted otherwise, all samples were annealed in Ar at 850 °C before and after plasma processing using the procedure detailed in **Methods**. The first annealing step is used to show that plasma processing is the dominant step responsible for emitter generation, while the subsequent annealing step stabilizes the emitters, as is discussed later. Red circles show the locations of emitters found in the crystals by a pixel-by-pixel analysis of spectra acquired from the PL maps. The plasma treatment elevates PL intensity preferentially at crystal edges and grain boundaries, which is where most of the SPEs discussed below are located in both as-prepared and plasma-processed samples. The preferential decoration of edges and extended defects by emitters is consistent with previous studies of hBN.[9] The as-prepared crystals contain very few emitters prior to etching, and they are all located at bright edge site regions, as is seen in Figure 1c. Plasma processing increased the number of emitters by a factor of three in this particular field of view, and some of the new emitters are present away from the edges of the flakes shown in Figure 1d.

Emitters in hBN are known to possess a broad range of zero phonon lines (ZPLs), attributed tentatively to variations in the structural compositions and charge states of color centers, local dielectric environments and strain.[9, 13, 16] Figure 1e therefore shows histograms of ZPL wavelengths for all emitters found in 5 different fields of view, the total area of crystals in each field is smaller than ~ 40 × 40 μm$^2$, before (8 emitters) and after (66 emitters) the Ar plasma treatment. Both histograms peak around 600 nm, but the plasma treatment yields emitters with a broader range of ZPL wavelengths and, notably, some emitters with wavelengths longer than 750 nm. A number of representative normalized spectra from emitters generated by the plasma treatment are shown in Figure 1f.

We proceed to characterize quantum optical properties of the emitters. Figure 2a shows the ZPLs of a representative emitter produced by the Ar plasma etch process obtained at room temperature and 11K, the spectra were fitted with a Lorentz function. The ZPLs maximum are located at ~ 711 nm and the full width at half maximum (FWHM) is ~ 16 nm at room temperature and ~ 2.8nm at 11K, respectively. In all our experiments, we did not observe a spectrometer limited FWHM of the emitters. The associated broadening at room temperature is likely caused by spectral diffusion or low energy phonon coupling.[17] To evaluate the excited state lifetime of the emitter, the PL decay time was measured using a 512 nm pulsed laser (Figure 2b) at room temperature, yielding an excited state lifetime of 2.4 ns after fit the curve with double-exponential. The lifetime is consistent with previous reports on single photon sources in hBN crystals.[7] The inset of figure 2b shows is the second order autocorrelation function ($g^{(2)}(\tau)$) obtained from this emitter. The dip at zero delay is well below 0.5 ($g^{(2)}(0) \sim 0.1$), confirming that the defect is a true single photon emitter.[7] The curve was fit with a three-level model, where $\tau_1$ and $\tau_2$ are excited and metastable state lifetimes, respectively.

$$g^{(2)}(\tau) = 1 - a\, e^{-\tau/\tau_1} + b\, e^{-\tau/\tau_2} \tag{1}$$

Overall, 10 out of the 66 emitters prepared with this method are single photon emitters. The rest showed very shallow dip indicating they are ensembles. Figure 2c depicts the fluorescence intensity as a function of excitation power. A standard three level system that has a ground, excited and a long lived metastable state was used to fit the data using the following equation,

$$I = I_\infty \frac{P}{P+P_{sat}} \tag{2}$$

where $I_\infty$ and $P_{max}$ are the emission rate and the excitation power at saturation, respectively. The saturation emission rate for this emitter is $5.4 \times 10^4$ at a saturation power of 910 µW. The brightness is comparable to previous reports of emitters in tape exfoliated hBN crystals.[13] Most emitters found after the Ar plasma etching were not stable and bleach in seconds. A post-etching annealing treatment was therefore performed to stabilize the emitters. This is demonstrated in Figure 2f which compares the stability of emitters measured before and after annealing (additional stability curves are shown in Figure S1). The same technique can be effectively used to pattern hBN flakes using selective masking to form emitter arrays. An example of such an array is shown in Figure S2.

To study the effects of plasma etching on hBN crystals, part of a hBN crystal was covered with photo-resist to protect the flake during plasma processing. Atomic force microscopy (AFM), Raman spectroscopy and X-Ray photoemission were then used to study the samples. Figure 3a shows an optical image of the hBN crystal obtained after etching. Red dashed lines indicate the boundary between the etched/protected areas, and the height difference is easily observed as a change in contrast in the image[18]. Figure 3b shows a height image of the sample surface obtained using AFM. The surface of the protected hBN remains very smooth (RMS roughness ~ 0.65 nm) which suggests it was not damaged during the etching process. By contrast, the etched area contains randomly distributed nano-islands (RMS roughness ~ 0.91 nm).[19-20] Furthermore, Ar plasma etching produces a ~ 2 nm step (Figure 3b and c) at the etched-protected interface, suggesting that the fabricated new emitters are located in the top four layers of the exfoliated hBN flake. The maximum penetration ranges of $Ar^+$ and $O^+$ ions and the corresponding vacancy generation depth profiles were simulated using the Monte Carlo code SRIM (stopping range of ions in matter).[21] The results are shown in figure 3d for an accelerating voltage of 400V (the self-biased voltage of the plasma system under the conditions used in this work, see **Methods** for details). Most of the ions penetrate 2 - 3nm below the hBN crystal surface, confirming that the plasma process is very shallow, and the emitters are fabricated within the top few nanometers of the surface. Argon and oxygen plasma treatments are both known to remove organic surface adsorbates and etch 2D layered materials with controlled etching rates.[22-28] The accelerated ions are also capable of breaking N-B bonds and creating N vacancies in the hBN lattice. As revealed by previous research[19-20, 22], N vacancies generated by Ar plasma etching are expected to be decorated with O atoms once exposed to air.

Figure 3e shows Raman spectra from the pristine and etched hBN regions of the crystal. A characteristic peak corresponding to the $E_{2g}$ phonon mode occurs at ~1395 $cm^{-1}$, in etched samples, the Raman peak is slightly shifted, typically to 1393 $cm^{-1}$. On the other hand, the FWHMs in the etched samples is smaller than in the pristine ones.[18] The red-shifted and sharpened Raman peaks and the rougher surface on the etched flake suggest that covalent bonds and strain in the crystals were altered, especially at the edge/grain boundaries.[18-20, 22, 29] To confirm the presence of the oxygen atoms in the hBN lattice, X-ray photoemission spectrum was performed on the Ar plasma treated samples as shown in figure 3f. The binding energy profile of B1s consists of 5 peaks located at 189.4, 190.5, 191.6, 192.6, 193.3eV. The peaks at 192.6 and 193.3eV suggests the formation of the B-N-O complex.[19-20, 30-31] Thus, we propose that the emitters generated by the Ar plasma treatment may be attributed to an oxygen related vacancy complex. Note that the emitters generated by an oxygen plasma (Figure S3) are different from the Ar plasma as the defects are oxidized simultaneously during the etch process.[23] Consequently, the emission from oxygen plasma treated hBN would originate from boron oxides, which are known to be bleach in seconds.[32]

To further confirm our hypothesis, we annealed hBN crystals in oxidative atmosphere under ambient pressure at 750°C for 30 minutes. At this temperature, pristine hBN crystals, even monolayers, are unlikely to be oxidized[33], regardless of thickness, but defects at grain boundaries may potentially be modified in the oxidizing atmosphere. Figure 4a shows a normalized PL intensity map obtained from a hBN crystal that was annealed in air. The positions of emitters found in this crystal are indicated by red circles, all of which are located at grain boundaries. AFM was used to study the morphology of this flake, as shown in figure 4b, adjacent flakes crumpled to form ridges at the grain boundary with a height of around 300 nm. We found that the emitters were mostly likely located at the ridges after correlate the PL mapping and the AFM image. The pristine defects in hBN are located at the edges of the flakes and the ZPLs range from 550nm to 650 nm, which is in consistent with previous reports.[9-12] In comparison, the new emitters are preferentially located at the ridges, or grain boundaries, with longer emission wavelength, which would be resulted from oxygen doping.

Figure 4c shows the PL spectra from a native defect and oxygen related defect on this crystal. The polar graph in figure 4d shows the emission from the new emitters are linearly polarized.

To confirm that oxygen indeed plays a role in generating the quantum emitters, first-principles calculations are performed on intrinsic vacancy defects and vacancies decorated with oxygen atoms using the density functional theory (DFT) method (See Figure S4-S6 in Supporting Information). Several defects, *i.e.* the anti-site vacancy, N-vacancy, B-vacancy and interlayer bonding, have previously been proposed to be responsible for the broad range of ZPL energies in hBN and evaluated using the Perdew-Burke-Ernzerhof (PBE) functional (See **Methods**).[7, 13] However, as it is widely accepted that PBE functional underestimates the electronic band gaps of semiconductors, we used hybrid functional (HSE06) to obtain more accurate electronic band structures (See details in Methods), and to evaluate a number of oxygen-related defects in hBN. Due to the presence of dangling bonds in the vacancies, the adsorption of O atoms is energetically favorable, resulting in a range of impurity states within the wide band gap of h-BN. Assuming that only spin-preserving transitions are allowed, we find that the B-vacancy defect saturated with two oxygen atoms ($V_BO_2$), as shown in Figure 5a, is the most likely candidate for color centers with longer wavelength emission. The calculated energy levels in Figure 5b show that there is a single transition from a potential ground state at 0.98 eV to a potential excited state located at 2.83 eV, resulting in a transition with an energy of 1.85 eV, which is very close to the experimental results. Note that quasi-particle approximation is not considered here due to extremely large amount of calculations in this system and thus this value is slightly larger than the measured values (a ZPL centered on 718 nm, corresponding to ~1.73 eV). Since this transition is optically polarized and occur between highly localized HOMO-LUMO states, therefore this defect is most likely to be the source of clean emission seen in experiment. All the other oxygen related defects do not have such atomic like states and defect levels are somehow mixed with the bulk bands therefore they are less likely to be responsible for the seen single photon emission.

**Conclusion**

In conclusion, we were able to engineer single emitters in hBN flakes by employing an argon plasma treatment. We showed that post-etching annealing process is required to stabilize the emitters, and the overall number of emitters increases by a factor of seven. AFM measurements confirmed that the new emitters are localized within the top few layers of hBN and X-Ray photoemission spectroscopy confirms the presence of oxygen atoms after plasma etching. Density functional theory elucidates on the possibility of oxygen related defects in hBN crystal matches well with our experimental results. While further measurements are needed to understand the origin of the emission, our results provide a compelling first evidence towards scalable engineering and patterning of room temperature quantum emitters in layered materials.

**Methods**

***Sample fabrication***: The hBN crystals were produced at atmospheric pressure using a molten mixture of nickel and chromium (weight ratio, 1:1) as the solvent, and isotopically enriched boron-10 metal as the boron source. The mixture was heated to and held at 1550 °C for 24 hours under flowing nitrogen. Slow (4°C/hour) cooling of this solution caused the hBN crystals to precipitate on the metal alloy surface.[34-35]

High quality hBN single crystals were mechanically exfoliated with scotch tape onto a Si substrate with a 300nm thermal oxide capping layer. The crystals were annealed at 850°C in Ar for 30 min after a cleaning process comprised of a 450°C anneal in air and a benign oxygen plasma treatment. These "as-prepared samples" were pre-characterised, and loaded into a vacuum chamber for Ar or $O_2$ plasma etching using a system equipped with a 13MHz radio-frequency (RF) plasma generator. The crystals were etched with a power of 200 W under a pressure of 180 moor for 2 mins at room temperature. Under these conditions the ions were accelerated by a DC self-bias of ~ 400V. The crystals were subsequently annealed at 850°C in Ar for 30mins to stabilize the emitters after plasma processing. The exfoliated hBN were also annealed at 750°C in air for 30mins, knowns as annealed in oxidative environment.

***Optical Characterization***: A home-built confocal setup was used for optical characterization of the crystals. Samples were excited with a 532 nm continuous wave (CW) laser (Gem 532™, Laser Quantum Ltd.) directed through a Glan-Taylor polarizer (Thorlabs Inc.) and a half waveplate, and focused onto the sample using a high numerical aperture (NA = 0.9, Nikon) objective lens. An X-Y piezo scanning mirror (FSM-300™) was used to perform confocal scanning. The excitation laser light was blocked with a 532 nm dichroic mirror (532 nm laser BrightLine™, Semrock) and a tilted 580nm long pass filter (Semrock) at the collection end. The PL at cryogenic temperature were collected by placing the hBN crystals in a home-built open-loop cryostat with flown liquid helium. The sample was excited with a computer-controlled continuous-wave (CW) Ti:Sap laser (SolsTis, M2 Inc.) The signal was then coupled into a graded index fiber and a fiber splitter was used to direct the light to a spectrometer (Acton SpectraPro™, Princeton Instrument Inc.) or to two avalanche photodiodes (Excelitas Technologies™) to collect spectra or autocorrelation data. PL mappings were gernerated using the photons collected from one of the APDs. Autocorrelation measurements were performed in a Hunbury Brown-Twiss configuration where the emission was equally separated and directed into two APDs. Lifetime measurements were performed using a 512 nm pulsed laser excitation source (PiL051X™, Advanced Laser Diode Systems GmbH) with a 100 ps pulse width and a 10 MHz repetition rate.

***Simulations***: All calculations are performed by using DFT method as implemented in the Vienna Ab initio Simulation Package (VASP).[36] The interaction between valence electrons and ions is described by the projector augmented wave (PAW) method, and (PBE) functional is chosen for the exchange and correlation interactions in geometric optimiztion.[37] Bearing in mind, the PBE always underestimates the energy band gaps of extended systems due to the delocalized description of electrons, the screened hybrid HSE06 functional is used to obtain more accurate electronic band structures by mixing semi-local and Hartree-Fock exchange, which has been widely tested in previous calculations.[38] A plane-wave basis set with a kinetic energy cutoff of 500 eV is used for all calculations. Pristine single-layer hBN was first geometry-optimized using the conventional cell and a 18 × 18 × 1 Monkhurst–Pack reciprocal space grid to an energy tolerance of 0.01 eV. A supercell containing 6 × 6 unit cells of $h$-BN and the vacuum thickness of 15 Å is adopted to avoid the interaction between two neighboring defects.

**Conflict of Interest:** The authors declare no competing financial interest.

**Supporting Information**

Supporting Information is available from the http://pubs.acs.org from the author.

**Acknowledgments**

The authors thank Dr. Luhua Li from Deakin University for the fruitful discussions. The hBN crystal growth at Kansas State University was supported by NSF grant CMMI 1538127. Financial support from the Australian Research Council (via DP140102721, DE130100592), FEI Company, and the Asian Office of Aerospace Research and Development grant FA2386-15-1-4044 are gratefully acknowledged. USTC group was supported by NSFC (21573204, 21421063), the MOST (2016YFA0200602), Strategic Priority Research Program of CAS (XDB01020300), and by USTC Supercomputer Centers.

**Figure Captions**

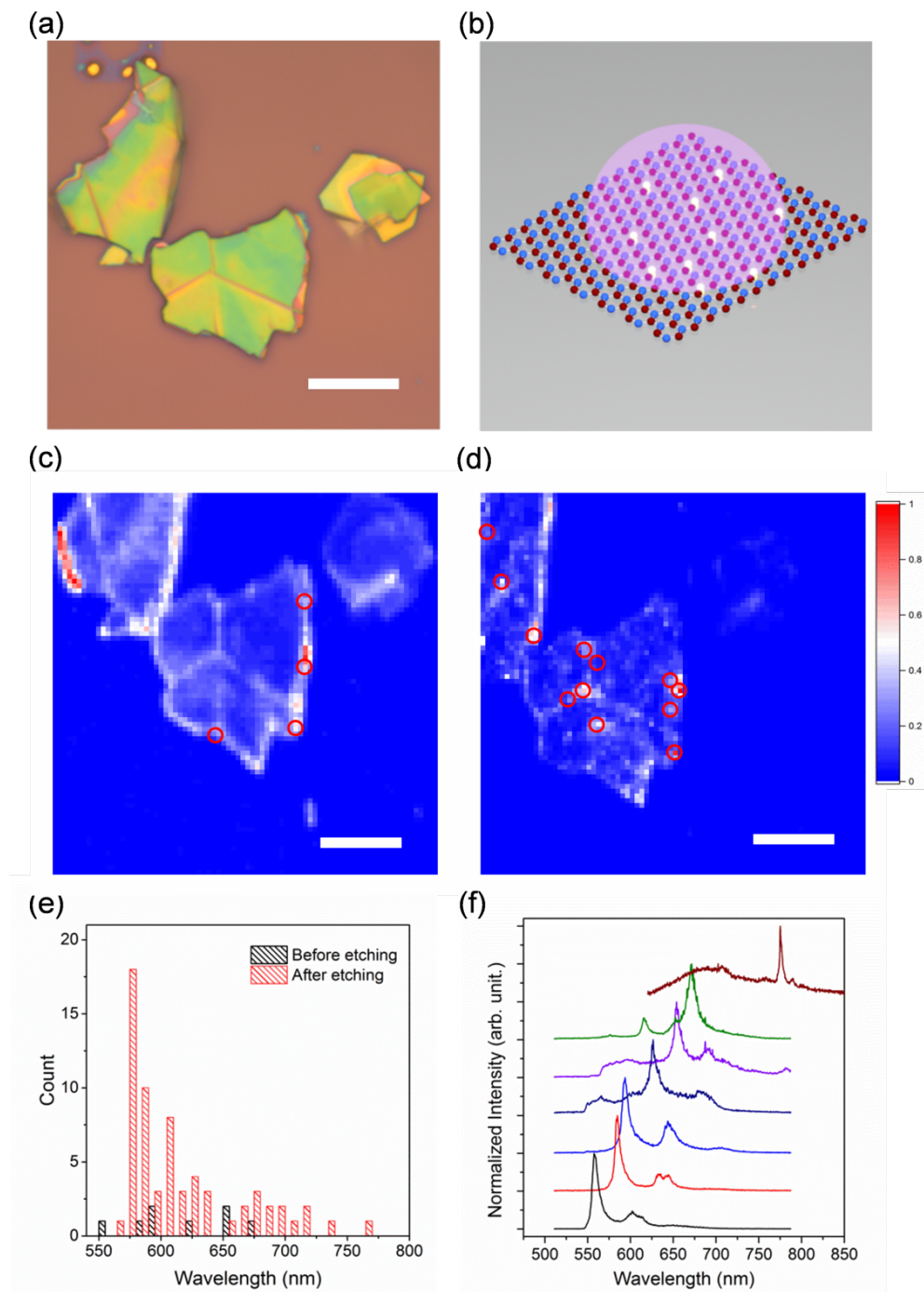

Figure 1 (a) Optical microscope image of hBN crystals. Scale bar: 10 μm. (b) Schematic image showing a hBN atomic sheet under plasma etching. (c,d) Confocal PL maps of the hBN crystals shown in (a) before and after Ar plasma etching. The red circles highlight the locations of emitters found before (c) and after (d) plasma etching. (e) Histograms of zero phonon line wavelengths of emitters found before and after plasma etching generated using a 5nm bin size. (f) Selected spectra of emitters found after plasma etching and annealing. The spectra are offset vertically for clarity

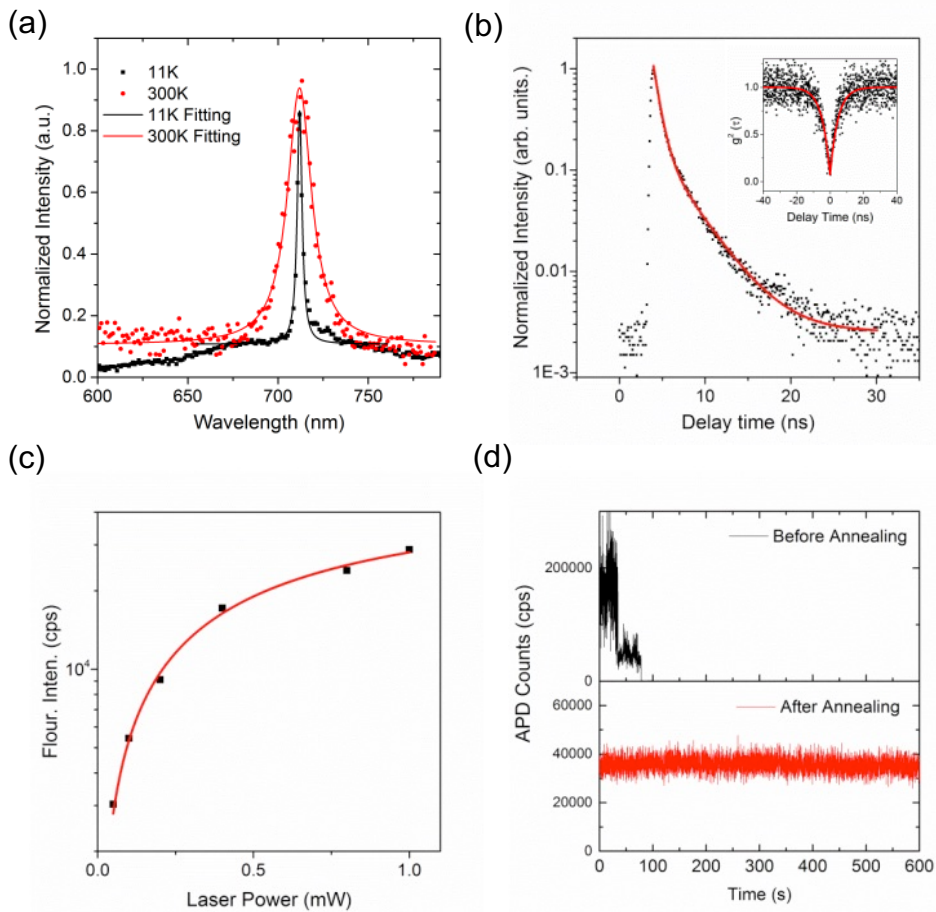

Figure 2 (a) A representative PL spectrum with an emission peak at 711 nm (fitted with a Lorentz function) obtained at room temperature and 11K from Ar plasma etched and annealed hBN crystals. (b) A time-resolved decay curve shows that the emitter lifetime is 2.4 ns. Inset: A second order autocorrelation function of the emitter fit using a three-level model with a $g^{(2)}(0) \sim 0.1$, confirming a single photon source. (c). A saturation curve of the emitter. (d) Representative emitter stability curves obtained before (black curve top panel) and after annealing (red curve bottom panel) of a plasma-etched hBN crystal.

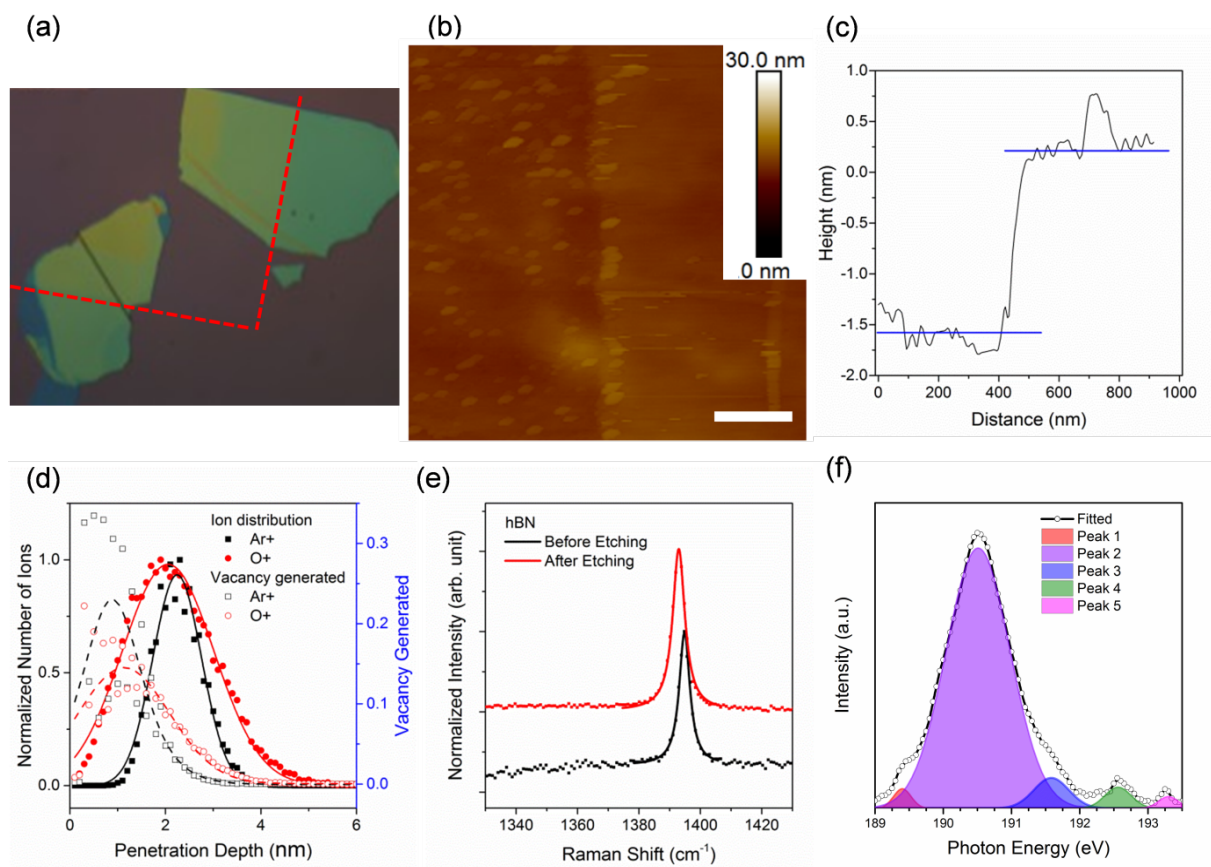

Figure 3 (a) Optical image of plasma-etched hBN crystals. The contrast seen at the dashed lines corresponds to the thickness difference produced by the etch treatment. (b) AFM image of the interface between etched (left) and protected (right) regions of a hBN crystal. Scale bar: 1 μm. (c) AFM height profile across the interface. (c) Ion depth distributions (filled squares and circles) and vacancy generation rates (hollow squares and circles) simulated for 400 eV $Ar^+$ and $O^+$ ions implanted into hBN. (e) Raman spectra obtained from pristine (black data points) and etched (red data points) regions of a hBN crystal. The spectra were offset vertically for clarity. The data were fit with Lorentizan functions. (f) The binding energy profiles for B1s in hBN crystals post plasma treatment.

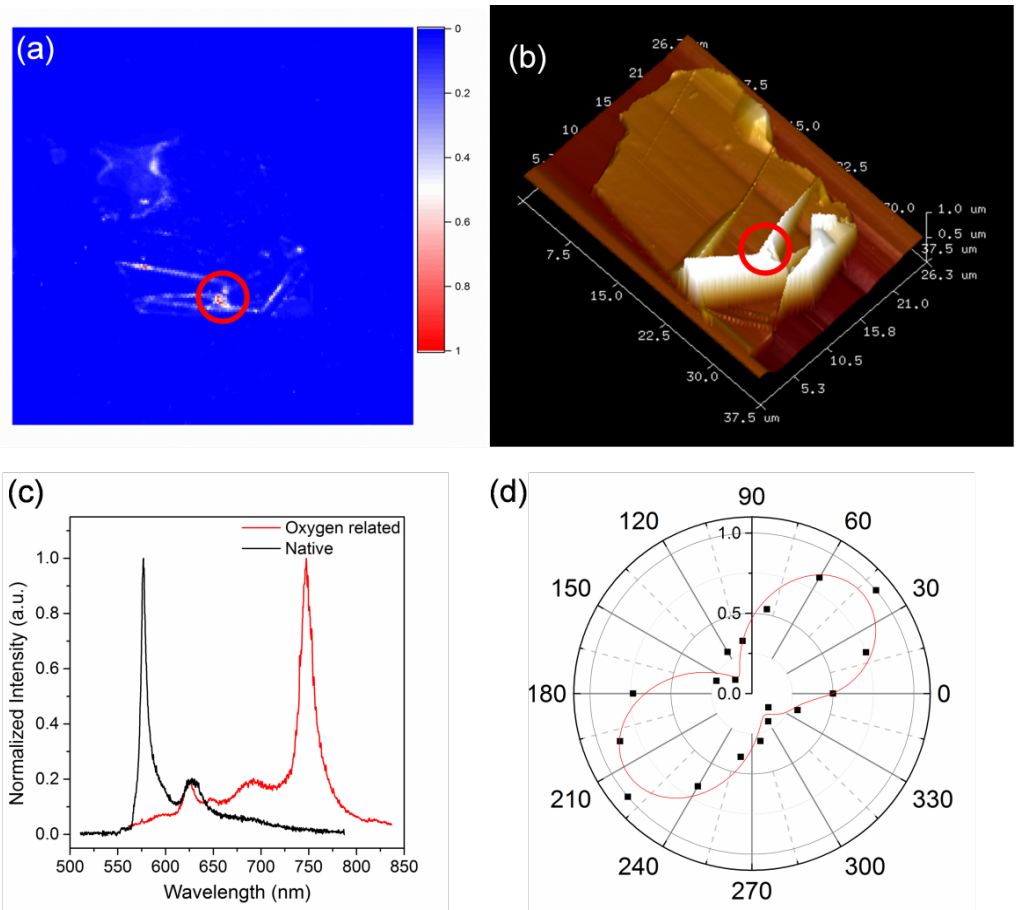

Figure 4 (a) Normalized confocal PL map taken from a hBN crystal annealed in oxidative environment at 750°C. The red circles show the locations of the emitters. (b) 3D AFM image of the same flake used in figure a. (c) Selective PL spectra excited with 532nm CW laser from native emission and oxygen related colour centres. (d) Polar graph of emission showing linearly polarized emission.

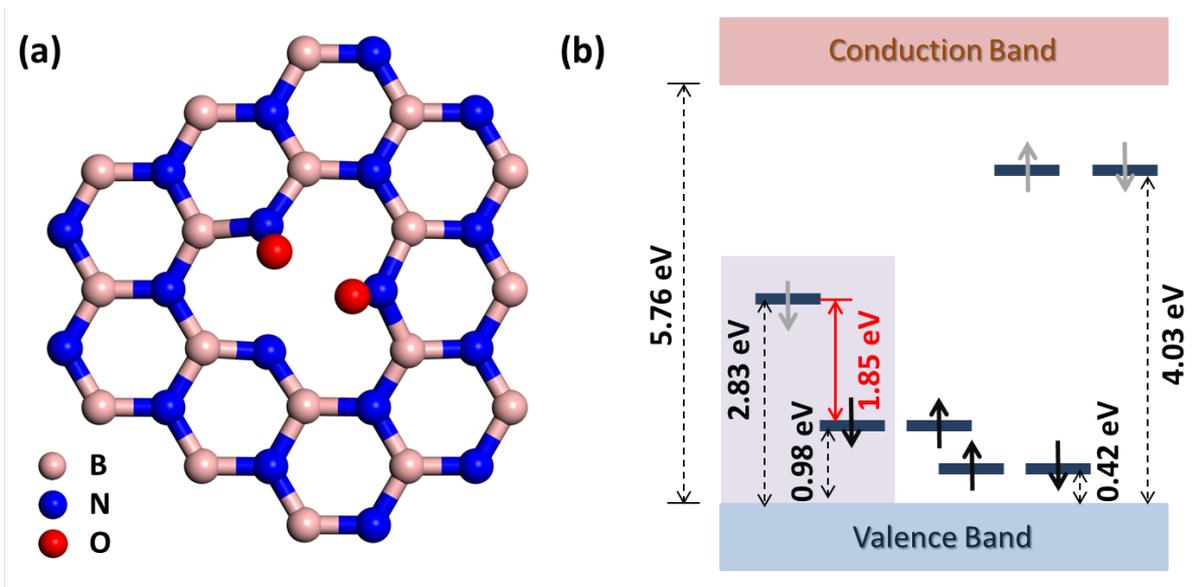

Figure 5. $V_BO_2$ defect model in the hBN lattice. (a) Schematic of the boron vacancy with oxygen atoms. (b) Simulated electronic structure using HSE06 functional. Black and grey arrows denote the occupied and unoccupied impurity states, respectively. Up and down arrows describe different spin. Other transitions from the unoccupied to occupied impurity states include: 2.41, 3.05, and 3.61 eV. Transition from unoccupied impurity states to VB include: 2.83 and 4.03 eV.

**Table of Contents**

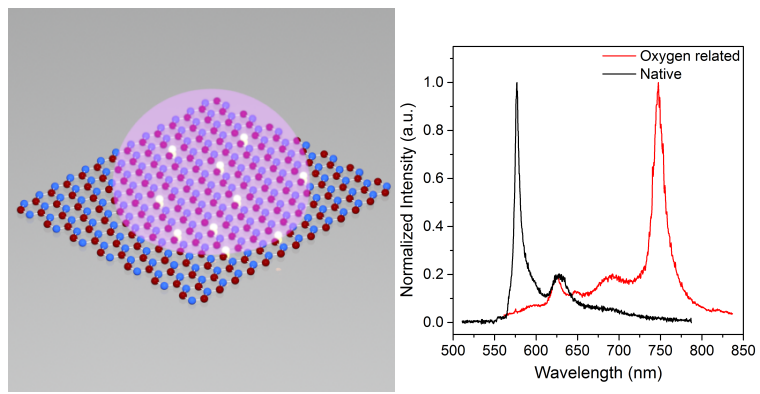